\documentclass[a4paper,twocolumn]{esapub} 

\usepackage{times}
\usepackage[square,numbers]{natbib}
\usepackage{graphicx}

\title{The GAIA Astrometric Survey of the Solar Neighborhood and its
Contribution to the Target Database for DARWIN/TPF}
\author[1,2,3]{A. Sozzetti}
\author[4]{S. Casertano}
\author[3]{M. G. Lattanzi}
\author[3]{A. Spagna}
\affil[1]{\it Harvard-Smithsonian Center for Astrophysics, 
Cambridge, MA 02138 USA, E-mail: asozzetti@cfa.harvard.edu} 
\affil[2]{\it University of Pittsburgh, Dept. of Physics \& Astronomy, 
Pittsburgh, PA 15260 USA} 
\affil[3]{\it Osservatorio Astronomico di Torino, 
10025 Pino Torinese, Italy, E-mail: lattanzi@to.astro.it, 
spagna@to.astro.it} 
\affil[4]{\it Space Telescope Science Institute, 
Baltimore, MD 21218 USA, E-mail: stefano@stsci.edu}

\begin{document}

\bibliographystyle{esa}


\maketitle

\begin{abstract}
We evaluate the potential of the ESA Cornerstone Mission GAIA in
helping populate the database of nearby stars ($d < 25$ pc) for
subsequent target selection for DARWIN/TPF. The GAIA high-precision
astrometric measurements will make it an ideal tool for a complete
screening of the expected several thousands stars within 25 pc in order
to identify and characterize (or rule out the presence of) Jupiter
signposts. GAIA astrometry will be instrumental in complementing
radial velocity surveys of F-G-K stars, and will more effectively
search for massive planets the large database of nearby M dwarfs, 
which are less easily accessible with precision spectroscopy. The ability
to determine the actual planet masses and inclination angles for
detected systems, especially those with low-mass primaries (M $<
0.6$ M$_\odot$), stems as a fundamental contribution GAIA will make
toward the final target selection for DARWIN/TPF, thus complementing 
exo-zodiacal dust emission observations from ground-based observatories 
such as Keck, LBTI, and VLTI. 
\end{abstract}

\section{Introduction}

The nearest stars to our Sun, within a conservative distance $d < 25$
pc, will constitute the ancillary database for which the
actual target list for DARWIN/TPF will be derived. The final
selection will be made on the base of characteristics such as
spectral type, age, metallicity, multiplicity, presence of
sub-stellar companions such as giant planets, and possible
evidence for rocky planets. It is remarkable how for the vast
majority of the stars in our immediate vicinity largely lacks 
such fundamental information. To date, about
2600 stars are known within 25 pc, but the total number, scaled
from the better-known sample within 5 pc, should be around 8000. 
So the database today is likely to be only 30\% complete, and data
for the presently known sample suffer from large uncertainties. In
particular, the database of low-luminosity M dwarfs lacks probably
about 80\% of the total population within 25 pc. A complete census
of this part of the sample of nearby stars has a special
connection to defining the DARWIN/TPF target list. Conventional wisdom
has it that M stars are not expected to have habitable rocky
planets because such planets would need to be so near their
primaries that tidal friction would halt or slow the planets' 
rotation, rendering them sterile. However, some research indicate that
a substantial atmosphere can redistribute heat around an otherwise
too-slowly rotating planet. Also, a typical M0 star has 1/2 the
mass of the Sun, and would heat a planet to terrestrial
temperatures at the position of Mercury's orbit where tidal
breaking might not be severe. Thus, we suggest that the large 
database of nearby M dwarfs should be kept under focus by future 
studies dedicated to the selection of the optimal target list for 
DARWIN/TPF.

\section{The Nearby Single Star Database}

\begin{figure*}[t]
\centering
\includegraphics[width=.9\linewidth]{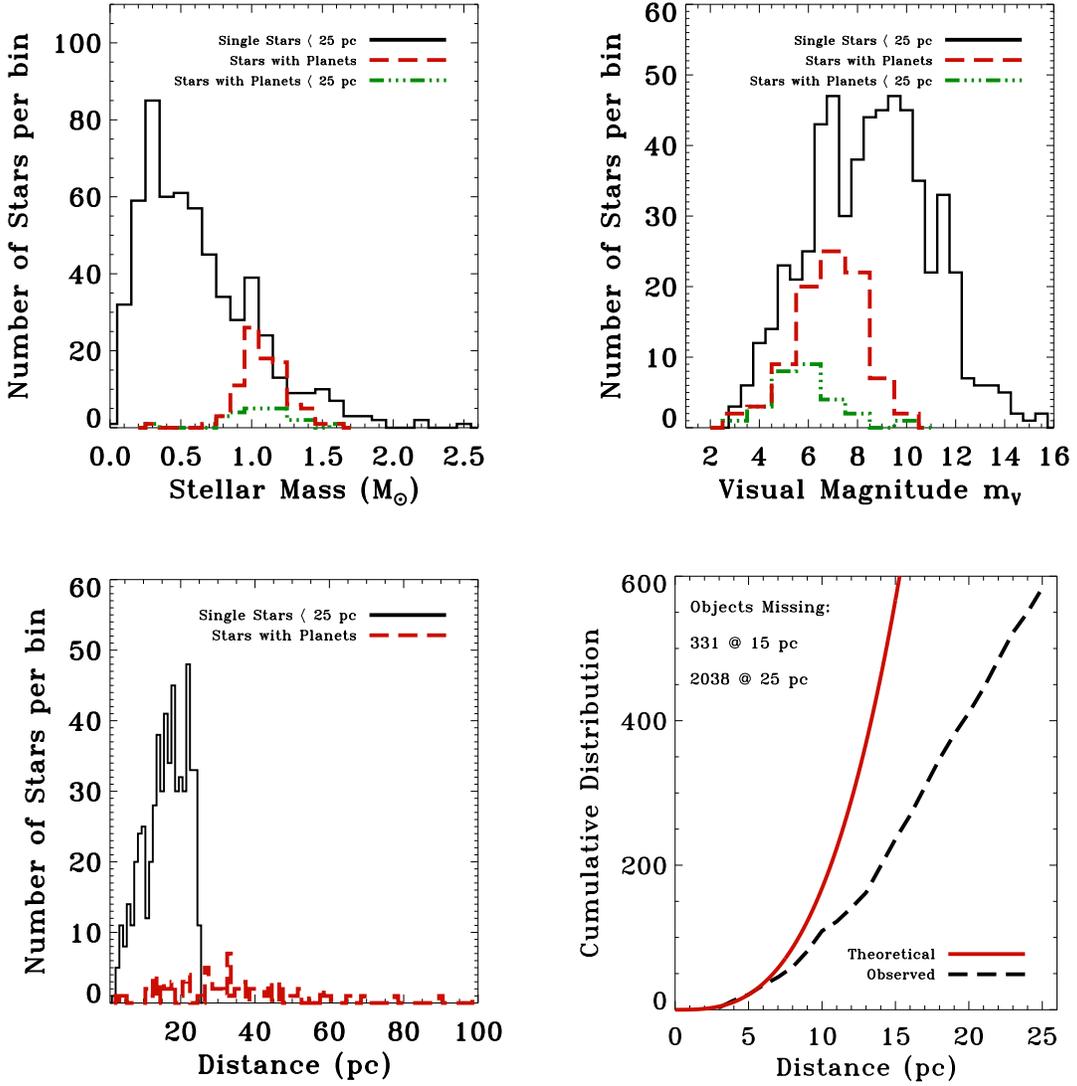}
\caption{Top left: distribution of nearby stars ($d < 25$ pc) as 
a function of mass, compared to the same distribution of stars 
with planets, and the subset of this sample within 25 pc; top right: 
the comparison made between the same distributions expressed as a 
function of apparent 
magnitude (for reference, 
$V = $ 14.3 for a 0.2-$M_\odot$ M5 dwarf at 25 pc); bottom left: 
comparison between the 
distributions of nearby stars and stars with planets, both expressed 
as a function of distance; bottom right: comparison between the 
theoretical and observed cumulative distributions of nearby stars) 
\label{fig1}}
\end{figure*}

The best targets for DARWIN/TPF are likely to be single stars.
Within a radius of 25 pc, 584 such stars are known today. Again
assuming the number of single stars scales with the volume, and 
extrapolating directly from the well-known sample of 21 such objects 
existing within 5 pc, one would conclude (lower right panel of 
Fig~\ref{fig1}) that $\sim 75\%$ of the reservoir of single stars is 
still missing. The radial velocity information about 
the existence of giant planets is available only for a fraction of the
known objects, the 100 or so moderately luminous F, G, and early K
type stars. The lack of knowledge for the majority of the lower
luminosity stars known today (and for all late K and M
dwarfs not identified yet as nearby stars) is clear when the
distributions of stellar objects in the solar neighborhood as a
function of V magnitude, mass, and distance are compared to the
analogous distributions for the planet-hosting stars, as shown in
the upper right and left and lower left panels of Fig.~\ref{fig1}.
Radial velocity surveys are reaching now completeness out to 50 pc for the
bright F-G stars ($V \leq 8$), and for planetary companions more
massive than Jupiter in the orbital radii range $< 3-4$ AU. Only a
limited number of K and M dwarfs has been monitored so far for
planets with comparable sensitivity to the one routinely reached
nowadays in the case of F and G dwarfs. Such objects will need 
careful studies in the coming years, as they will likely
constitute the majority of the list of targets for DARWIN/TPF. To
this end, information from radial velocity measurements (including
spectroscopic orbits for detected planetary mass companions) will
have to be complemented by precision astrometry.

\section{Target List Selection Criteria}

\begin{figure*}[t]
\centering
\includegraphics[width=.92\linewidth]{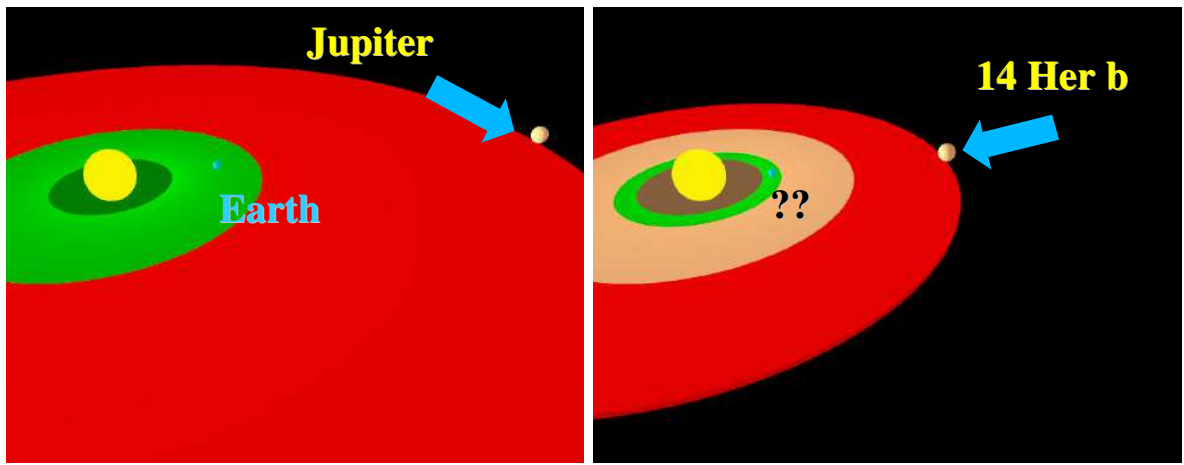}
\caption{Left: the Habitable Zone (green) and Exclusion Zone (red) in 
our Solar System. Right: for 14 Her, harboring a gas giant planet 
with a minimum mass of $\sim$ 3.3 M$_J$ orbiting with 
a period of 4.4 yr, the formation of a terrestrial planet could still have 
occurred in principle in a narrow region inside the Habitable Zone 
\label{fig2}}
\end{figure*}

Once a database of stars has been searched based on spectral type,
metallicity, age, and distance information, the criteria to be
adopted for refining the list of possible targets focus on their
Habitable Zones and Exclusion Zones. The Habitable Zone~\cite{kasting93} 
is conventionally defined by the distance from a given star at which
the temperature is such that water can be present in the liquid
phase. The center of the Habitable Zone (whose distance depends on
the mass of the parent star M$_\star$) can be roughly identified by the
formula:
\begin{equation}
\frac{\displaystyle T_{HZ}}{\displaystyle T_\oplus} =
\left(\frac{\displaystyle M_\star}{\displaystyle
M_\odot}\right)^{7/4},
\label{hz}
\end{equation}
where $T_\oplus$ is the Earth's orbital period. 
The inner and outer boundaries of the Habitable Zone for main
sequence dwarfs are placed at roughly $T_{HZ,in} \simeq 0.7\times
T_{HZ}$ and $T_{HZ,out} \simeq 2\times T_{HZ}$, respectively. The
Exclusion Zone~\cite{wetherill96} is operationally defined 
by the dynamical constraint:
\begin{equation}
T_G > 6\times T_R,
\end{equation}
which states that for a rocky planet to form in the Habitable Zone
of a star then a giant planet must form on an orbital period $T_G$
at least six times larger than the period $T_R$ of the rocky
planet. This constitutes a more stringent constraint than the one
based on the dynamical stability analysis of the orbits in the
Habitable Zone of a star already hosting a gas giant, as this
approach does not take into account the possibility that such
rocky planets may have actually formed in these systems, at such
privileged distances, in the first place. As of today, only one
star, among the F and G type objects within 25 pc from the Sun, is
known to harbor a giant planet at an orbital radius such that 
in principle the formation of
a rocky object in its Habitable Zone might not have been
prevented: this is the bright ($V = 6.61$), nearby ($d = 18.15$ pc) G
dwarf 14 Herculis (14 Her). Fig.~\ref{fig2} summarizes the concepts
of Habitable and Exclusion Zones as they have been realized in our
solar system and in the 14 Her planet-star system.

In light of these constraints, a given star may then be selected
as a DARWIN/TPF target if one of the three following conditions
apply:

\begin{itemize}
\item[1)] A rocky planet in its Habitable Zone has been already
detected by other means (the highest priority targets);

\item[2)] A `Jupiter signpost' has been detected at an orbital radius from
the star such that it has not precluded the possible formation of
Earth-sized bodies in the region where water is liquid;

\item[3)] Close scrutiny of the star has not produced any detection
of giant planets for a wide range of orbital radii, thus
planetesimal accumulation in the inner regions has been in
principle free to occur.
\end{itemize}

Today, the information needed to make these selections 
is available in part for only a limited number of
F-G type stars within 25 pc, thanks to precision (3-5 m/s)
ground-based radial velocity measurements. As discussed above,
only 1 star within 25 pc falls into the second class, while, due 
to insufficient sensitivity of present-day instrumentation, no 
objects have been found that fulfill criterion 1). The need to
extend the surveys to late-type objects and lower planet masses is
thus clear.

\section{How Many Targets for DARWIN/TPF?}

\begin{figure*}[t]
\centering
\includegraphics[width=1.\linewidth]{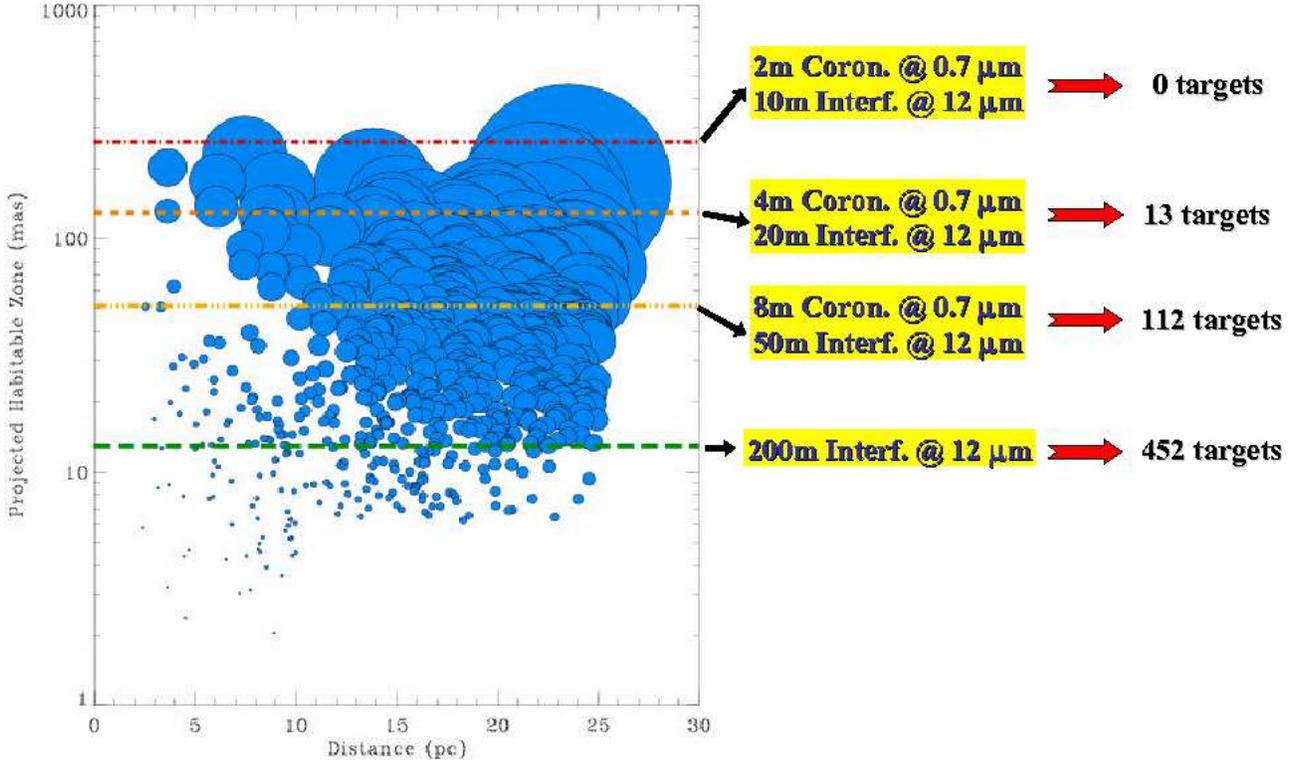}
\caption{Angular scale of the Habitable Zone around known single stars 
within 25 pc. Size and location of the Habitable Zone vary according 
to stellar mass, as per Eq.~\ref{hz} (M stars have the smallest 
circles, at the shortest distances from their parent star). Lines of 
different colors identify the angular resolution (defined as 
$\lambda/B$ and 3.6$\times\lambda/D$, where $B$ and $D$ are the 
interferometric baseline and monolithic telescope diameter, 
respectively) for various types of interferometric/coronagraphic 
configurations proposed for DARWIN/TPF\label{fig3}}
\end{figure*}

Beside a number of technological challenges, today the
most important source of uncertainty for DARWIN/TPF from a
scientific point of view is the fact that the frequency
of terrestrial planets $\eta_\oplus$ is completely unknown. Naive
extrapolations based on the current database of gas giants would
be arbitrary at best. During the next 5-10 years transit
photometry missions from space (Kepler, Eddington) will have the
sensitivity to detect Earth-sized objects transiting on the disk
of their parent stars, thus it will become possible to derive a
first estimate of $\eta_\oplus$. However, these results will not
be conclusive, and a scientifically meaningful search for
habitable planets with DARWIN/TPF should be designed so that a
null result (no habitable planets found) be statistically
significant. In other words, DARWIN/TPF should examine enough
stars, with enough sensitivity and operational lifetime, so that
we can draw valid statistical conclusions about the prevalence and
properties of rocky planets found in other solar systems.

The two architectures for DARWIN/TPF presently under study by 
the National Aeronautics and Space Administration (NASA) and by the 
European Space Agency (ESA) are $a)$ an IR nulling interferometer 
operating either on a fixed structure or in a separated spacecraft 
configuration, and $b)$ a monolithic telescope with a coronagraph 
and/or apodized aperture operating at visible wavelengths~\cite{summary02}. 
Fig.~\ref{fig3} shows the projected location and width of the 
Habitable Zone (in milli-arcseconds) for the 584 stars known 
within 25 pc from the Sun. For comparison, the limiting resolutions 
of the abovementioned architectures are over-plotted, for a variety 
of values of the monolithic telescope diameter $D$ or the 
interferometric baseline $B$. As a general result, in order to sample 
{\it at least} a few hundred targets, telescopes in the 
visible with $D\geq 10$m or interferometers with $B\geq 100$m 
will be required. If the selection of targets was limited only to the 
few tens or so of luminous F-G-K dwarfs, the failure to find 
habitable planets could raise questions about the relevance of 
the sample, not about the frequency of terrestrial planets. 
For this reason, the final target list for DARWIN/TPF could be 
broadened to include a significant number of M dwarfs. Only a few 
of them have been searched so far for planets, and many of them 
have not been discovered yet as nearby stars to our Sun. 
The GAIA all-sky astrometric survey has the potential to crucially 
help filling this important gap in our knowledge of the database 
of nearby stars to our Sun.

\section{The Role of the GAIA Survey}
\begin{figure*}[t]
\centering
\includegraphics[width=.9\linewidth]{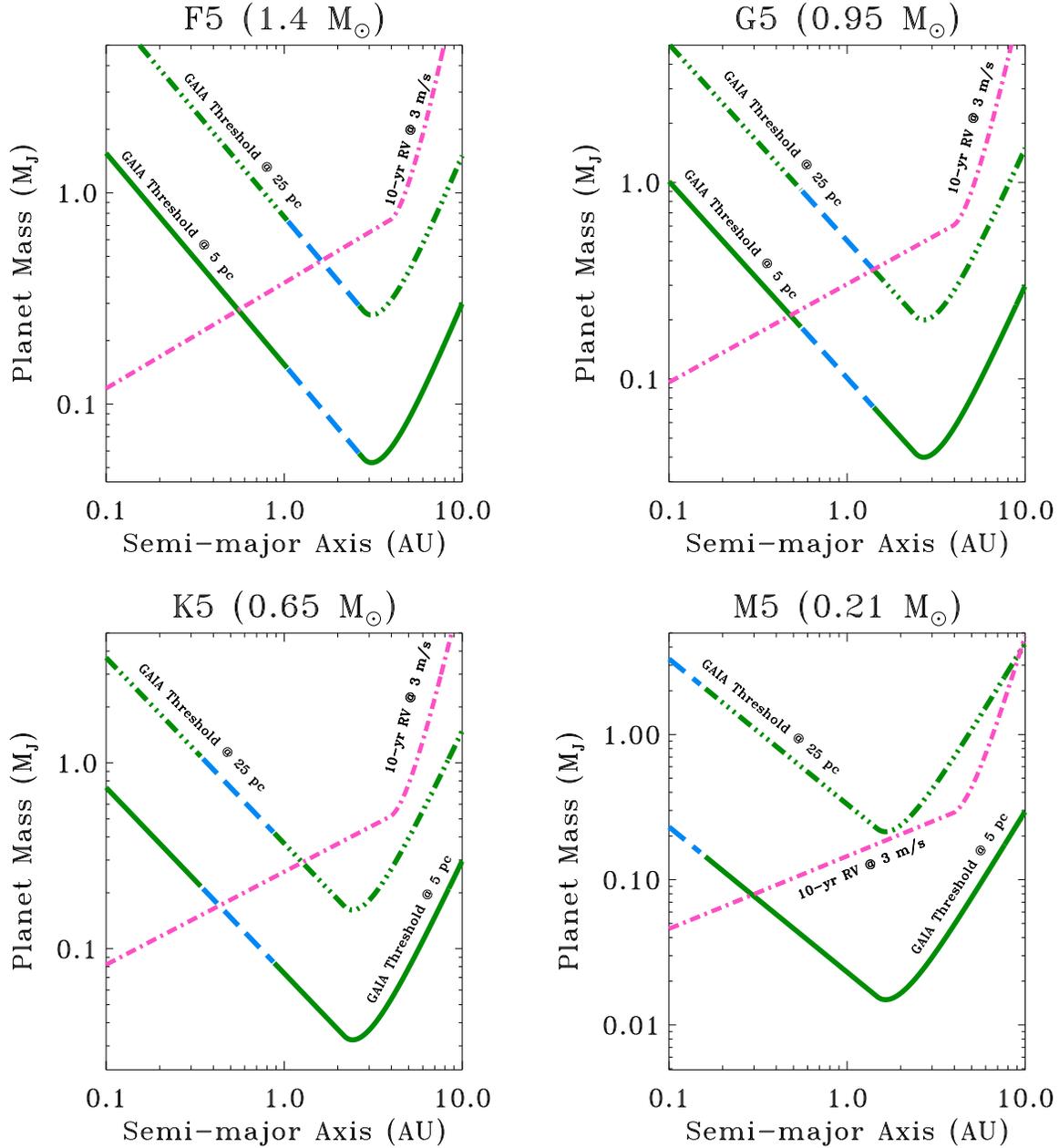}
\caption{GAIA planet discovery space as a function or orbital 
radius, stellar spectral type (F-G-K-M, from top left to bottom right), 
and distance from the observer (5 and 25 pc, solid and dashed-dotted-dotted 
green lines, respectively). For reference, the blue dotted segment 
identifies the star's Habitable Zone. The dashed-dotted purple 
lines identify the planet discovery space for 3 m/s precision 
radial velocity measurements. \label{fig4}}
\end{figure*}

The size of the stellar sample out to 150-200 pc to be
investigated for planets (hundreds of thousands of objects)
constitutes the most significant contribution GAIA will provide to
the science of extra-solar planets (e.g.,~\cite{lattanzi00b},
~\cite{sozzetti01}). 
Table~\ref{nplan} shows how, given reasonable 
assumptions on the planetary frequency as a function of orbital 
radius, on the detection threshold, and on the accuracy of orbit 
determination, GAIA will be capable of discovering 
thousands of planets around relatively nearby main-sequence stars, 
and it will accurately measure the orbital characteristics and 
actual masses for a significant fraction of the detected systems
\cite{lattanzi00a}~. 

\begin{table}[h!]
\begin{center}
\caption{(1) Number $N_\mathrm{d}$ of 
giant planets that could be detected by 
GAIA (at the 95\% confidence level), as function of 
increasing distance from the Sun. 
(2) Fraction $N_\mathrm{m}$ 
of detected planets for which orbital elements and masses can be 
accurately measured 
(to better than 20\%). A uniform frequency distribution of 1.3\%
planets per 1-AU bin is assumed}
   \renewcommand{\arraystretch}{1.4}
   \setlength\tabcolsep{6pt}
   \begin{tabular}{ccccc}
	\noalign{\smallskip}
	\noalign{\smallskip}
       \hline\noalign{\smallskip}
       $\Delta d$ (pc) & $N_\star$  & $\Delta a$ (AU)
&  $N_{\rm d}$ (1) & $N_\mathrm{m}$ (2)\\
       \noalign{\smallskip}
       \hline
       \noalign{\smallskip}
0-100 & $\sim$61\,000 & 1.3 - 5.3 & $\geq 1600$ & $\geq 640$ \\ 
100-150 & $\sim$114\,000 & 1.8 - 3.9 & $\geq 1600$ & $\geq 750$ \\
150-200 & $\sim$295\,000 & 2.5 - 3.3 & $\geq 1500$ & $\geq 750$ \\

      \hline\noalign{\smallskip}
   \end{tabular}
\label{nplan}
\end{center}
\end{table}
Thus, the results derived from GAIA 
high-precision astrometric measurements will help decisively 
improve our understanding of orbital parameters and actual mass
distributions, and they will provide important data to determine
the correct theoretical models of formation, migration, and
dynamical evolution.

With the current payload design~\cite{perryman01}, the range
of planetary masses between 1 Earth-mass and a few Earth-masses
will only be marginally accessible to GAIA's all-sky survey. Its
astrometric accuracy will be sufficient to address the issue of
their existence only around a handful of the closest stars, within
a few pc from the Sun. The most tantalizing targets for
DARWIN/TPF, those for which Earth-sized bodies in their Habitable 
Zones have already been detected, are thus likely to be 
found by SIM (see for example~\cite{sozzetti02}). Nevertheless,
GAIA's contribute to the search for rocky, possibly habitable
planets will be significant. In particular, GAIA has the potential
to complement radial velocity measurements in order to identify
good targets for DARWIN/TPF responding to selection criteria 2)
and 3). With an expected single-measurement precision of 8-10 $\mu$as
at V=13 or brighter, GAIA will be capable of monitoring the vast
majority of the nearby stars within 25 pc with sufficient accuracy
to confirm Jupiter signposts detections by radial velocity
techniques. In particular, GAIA will extend the spectroscopic
surveys at the faint end (late K through M dwarfs), where the size
of the sample and detectable orbital period ranges from ground
will likely be limited by telescope time constraints. 

For any particular star, the boundary of discovery space for astrometry 
and radial velocity---the 
dividing line between detectable and non-detectable---is found by
equating the minimum signal (the astrometric signature or the radial 
velocity semi-amplitude, respectively) required for discovery to the actual 
magnitude of the gravitational pull induced by the planet on its parent 
star~\cite{sozzetti02}. For a given
stellar mass (and distance from the observer in the astrometric case), 
it is then possible to
determine the minimum detectable mass M$_{\mathrm{p,min}}$ as a
function of the semi-major axis of the planetary orbit
$a_{\mathrm{p}}$. To take into account the loss in sensitivity for
periods $T$ longer than the mission length $L$ (set to 5 yr for GAIA, and 
10 yr for ground-based radial velocity surveys), we have
parameterized M$_{\mathrm{p,min}}$ in the two cases as follows:

\footnotesize
\[
\mathrm{M}^{(A)}_{\mathrm{p,min}} = \left\{\begin{array}{ll}
\kappa\times\sigma_{\mathrm{A}}
\times \frac{\displaystyle M_\star d}{\displaystyle a_\mathrm{p}},
 & \mbox{$T \leq \lambda L$}\\
\kappa\times\sigma_{\mathrm{A}}\times\frac{\displaystyle M_\star d}
{\displaystyle a_\mathrm{p}}\times
\sin^{-2}\left(\frac{\displaystyle \pi\lambda L}{\displaystyle 2T}\right),
& \mbox{$T > \lambda L$}
\end{array}
\right.
\]
\normalsize

\footnotesize
\[
\mathrm{M}^{(RV)}_{\mathrm{p,min}} = \left\{\begin{array}{l}
\kappa\times\sigma_{\mathrm{RV}}
\times \sqrt{M_\star a_\mathrm{p}}\,,\hskip 2.44cm \mbox{$T \leq \lambda L$} \\
\kappa\times\sigma_{\mathrm{RV}}\times\sqrt{M_\star a_\mathrm{p}}\times
\sin^{-2.4}\left(\frac{\displaystyle \pi\lambda L}{\displaystyle 2T}
\right), \,\mbox{$T > \lambda L$}
\end{array}
\right.
\]
\normalsize

In the plane defined by the mass of the planet M$_\mathrm{p}$ and
the orbital semi-major axis $a_\mathrm{p}$, these parametric
equations identify families of curves with the same shape, having
an absolute minimum at the value of $a_\mathrm{p}$ corresponding
to a period in years equal to the fraction $\lambda$ of the 
time-span of the observations where the sensitivity is greatest.

Fig.~\ref{fig4} shows the minimum detectable planet mass (M$_\mathrm{J}$) 
by GAIA as a function of orbital radius, stellar mass, and distance 
from the Sun (assuming a single-measurement error 
$\sigma_A = 10$ $\mu$as on the one-dimensional 
coordinate measured by GAIA along the scan direction). 
For comparison, the equivalent detection curves are also plotted 
for ground-based radial velocity surveys with an adopted 
single-measurement precision $\sigma_\mathrm{RV} = 3$ m/s, and assuming 
that integration times can be modified in order to reach the same 
sensitivity on stars of different spectral type. For both techniques, 
a signal three times larger ($\kappa = 3$) than the measurement precision 
is required for secure detection. For orbital periods exceeding 
the time-span of observations ($T > \lambda L$, with $\lambda = 7/8$) 
the loss in 
sensitivity (i.e. increasingly larger values of M$_\mathrm{p,min}$ 
needed for 
detection) for spectroscopy is faster than for astrometry, as the 
radial velocity amplitude decreases as the orbit gets larger. 

The GAIA
astrometric survey will provide, for all detected giant planets
for which an  orbital solution can be obtained, two fundamental
parameters which are needed in order to probe the actual validity
of a potential DARWIN/TPF target. First of all, it will derive
meaningful estimates of the actual mass of a planet, allowing
theorists to establish whether  dynamical interactions may have
prevented a rocky planet in the Habitable Zone from forming.
Second, it will measure the inclination of the orbital plane, thus
complementing studies of exo-zodiacal cloud emission from the
system under investigation (with ground-based facilities such as
Keck, VLTI, and LBTI), and allowing in turn to carefully select
possible targets in which the orbital inclination is not too close
to edge on (as the presence of zodiacal emission in systems with
inclination angles $i > 60^\circ$ may hamper the planet finding
capabilities of DARWIN/TPF).

\section{Conclusions}

Precision astrometry during the next decade will begin
complementing effectively planet searches based on radial velocity
measurements. In its global astrometric survey, GAIA will search
for the presence of massive planets all expected stars within 25 pc from
the Sun, including the large database of M dwarfs. For all
potentially interesting systems for DARWIN/TPF according to
selection criteria 2) and 3), GAIA has the potential to refine the
selected lists by providing meaningful estimates of two
fundamental parameters, actual planet mass and orbital
inclination, or by ruling out the presence of massive objects that
may have hampered rocky planet formation. The data GAIA will
provide on the presence (or absence) of Jupiter signposts orbiting
the faint end of the stellar sample in the solar neighborhood will
constitute very valuable knowledge that will crucially complement
the information coming from other techniques at the moment of the
final selection of targets for DARWIN/TPF.

\section*{Acknowledgements}

The authors gratefully acknowledge partial financial support from the
Italian Space Agency under Contract ASI-I/R/117/01. This 
research has made use of the NStars Database 
(http://nstars.arc.nasa.gov/).


\begin{thebibliography}{15}
\bibitem{kasting93}
$1.\;$ Kasting  J. F., Whitmire D. P., Reynolds R. T., Habitable 
Zones Around Main Sequence Stars, {\it Icarus}, 101, 108, 1993
\bibitem{lattanzi00a}
$2.\;$ Lattanzi M.G., Sozzetti A., Spagna A., Extra-Solar Planets 
with GAIA, in {\it From Extra-solar
Planets to Cosmology: The VLT Opening Symposium}, ed. J. Bergeron
\& A. Renzini (Berlin: Springer-Verlag), 479, 2000a
\bibitem{lattanzi00b}
$3.\;$ Lattanzi M. G., Spagna A., Sozzetti A., Casertano S., 
Space-Borne Global Astrometric Surveys: 
the Hunt for Extrasolar Planets, {\it MNRAS}, 317, 211, 2000b
\bibitem{perryman01}
$4.\;$ Perryman M. A. C., et al., GAIA: 
Composition, Formation and Evolution of the Galaxy, {\it A\&A}, 
369, 339, 2001
\bibitem{sozzetti01}
$5.\;$ Sozzetti A., Casertano S., Lattanzi M. G., Spagna A., 
Detection and Measurement of Planetary Systems with GAIA, {\it A\&A},
373, L21, 2001
\bibitem{sozzetti02}
$6.\;$ Sozzetti A., Casertano S., Brown R. A., Lattanzi M. G., 
Narrow-Angle Astrometry with the Space Interferometry Mission: 
The Search for Extrasolar Planets. 
I. Detection and Characterization of Single Planets, {\it PASP}, 
114, 1173, 2002
\bibitem{summary02}
$7.\;$ Summary Report on Architecture Studies for the Terrestrial 
Planet Finder, ed. C. A. Beichman, C. R. Coulter, C. A. Lindensmith \& 
P. R. Lawson, {\it JPL Publications}, 02-011
\bibitem{wetherill96}
$8.\;$ Wetherill G. W., The Formation and Habitability of 
Extra-Solar Planets, {\it Icarus}, 119, 219, 1996




\end{thebibliography}
\end{document}